\documentclass[useAMS,usenatbib]{mn2e}

\usepackage{psfig, epsf, epsfig}

\title[Feedback effects of aspherical supernovae]{Feedback effects
of aspherical supernovae explosions on galaxies}
\author[K. Bekki, T. ]
{Kenji Bekki${}^1$\thanks{E-mail:
bekki@cyllene.uwa.edu.au},
Toshikazu Shigeyama${}^2$,
and Takuji  Tsujimoto${}^3$ \\
${}^1$ICRAR M468
The University of Western Australia
35 Stirling Hwy, Crawley
Western Australia, 6009 \\
${}^2$
Research Center for the Early Universe, Graduate School 
of Science, University of Tokyo,
Bunkyo-ku, Tokyo, 113-0033 \\
${}^3$
National Astronomical Observatory, Mitaka-shi, Tokyo 181-8588, Japan \\}

\begin{document}

\date{Accepted, Received 2005 February 20; in original form }

\pagerange{\pageref{firstpage}--\pageref{lastpage}} \pubyear{2005}

\maketitle

\label{firstpage}

\begin{abstract}

We investigate how explosions of aspherical supernovae (A-SNe)
can influence star formation histories and chemical evolution of dwarf
galaxies by using a new chemodynamical model. We mainly present 
the numerical results of two comparative models so that the
A-SN feedback effects on galaxies 
can be more clearly seen.
SNe originating from stars with masses larger than $30 {\rm M}_{\odot}$
are A-SNe in the ``ASN'' model
whereas all SNe are spherical ones (S-SNe)
in the ``SSN'' model.
Each S-SN and A-SN are assumed to release feedback energy 
of $10^{51}$ erg and $10^{52}$ erg, respectively,
and chemical yields and feedback energy of A-SN ejecta depend on angles between
the axis of symmetry and the ejection directions.
We find that star formation can become at least by a factor of $\sim 3$ 
lower in the ASN model in comparison with the SSN one owing to the
more energetic feedback of A-SNe.
As a result of  this,
chemical evolution can proceed very slowly in the ASN model.
A-SN feedback effects can play a significant role in the formation
of giant gaseous holes and energetic gaseous outflow 
and unique chemical abundances (e.g., high [Mg/Ca]).
Based on these results, we provide a number of implications of
the A-SN feedback effects on galaxy formation and evolution.
\end{abstract}


\begin{table*}
\centering
\begin{minipage}{175mm}
\caption{Description of the basic parameter values
and some results for representative models.}
\begin{tabular}{ccccccc}
{Model name}
& {$M_{\rm dm,dw}$
\footnote{The initial total  mass of dark matter halo
in a dwarf  in units of $10^8 {\rm M}_{\odot}$.}}
& {$M_{\rm s,dw}$
\footnote{The initial total  mass of stellar disk 
in a dwarf  in units of $10^8 {\rm M}_{\odot}$.}}
& {$M_{\rm g, dw}$
\footnote{The initial total  mass of  gaseous disk
in a dwarf  in units of $10^8 {\rm M}_{\odot}$.}}
& {SN-type for $8 \le m_{\rm sn}/M_{\odot}  \le 30$ 
\footnote{The mass of each SN is denoted as $m_{\rm sn}$
and A-SN and S-SN represent aspherical and spherical SNe,
respectively.}}
& SN-type for $30 < m_{\rm sn}/M_{\odot}$  
& {$\epsilon_{\rm sf}$
\footnote{The star formation efficiency defined as 
$M_{\rm ns, dw}/M_{\rm g, dw}$, where $M_{\rm ns, dw}$ is
the total mass of new star formed until $T=1.4$ Gyr.}} \\
ASN & 20.0 & 1.0 & 1.0 & S-SN & A-SN & $9.5 \times 10^{-3}$ \\
SSN & 20.0 & 1.0 & 1.0 & S-SN & S-SN & $1.1 \times 10^{-1}$ \\
\end{tabular}
\end{minipage}
\end{table*}

\section{Introduction}

Feedback effects of supernova (SN) explosions  have long been considered to be
one of key determinants in galaxy formation and evolution
(e.g., Larson 1974; Dekel \& Silk 1986). 
SN feedback effects depending on galactic properties
(e.g., masses and potential depth)
have been discussed extensively in variously different
contexts of galaxy formation, such as the origin of the color-magnitude
diagrams in elliptical galaxies (e.g., Arimoto \& Yoshii 1980),
the formation of galactic disks  (e.g., Navarro \& White 1993),
global mass loss in low-mass galaxies  (e.g., Dekel \& Silk 1986),
and the formation of cored dark matter halos in dwarfs 
(e.g., Governato et al. 2010). 
Chemodynamical simulations have played significant roles  
in better understanding the SN feedback effects on chemical and dynamical
evolution of galaxies (e.g., Theis et al. 1992; 
Kawata \& Gibson 2003; Revaz \& Jablonka 2012).

A growing number of observational studies on host galaxies
for gamma-ray bursts (GRBs) have revealed
that most of the
GRB hosts at relatively low redshifts ($z<2$)
are  star-forming,  metal-poor, and dwarf-like galaxies
(e.g., Savaglio 2008; Savaglio et al. 2009).
More energetic explosion events
of jet-induced  supernovae are considered to be
associated with the origin of long-duration GRBs
(e.g., Woosley \& Bloom 2003) and
such supernovae are extreme cases of aspherical
supernovae (``A-SNe'').
Therefore,
the detection of a GRB event in a galaxy implies
that the galaxy has recently experienced
a  much larger number of A-SN events.
Given that A-SNe can release a significantly more amount
of energy ($\sim 10^{52}$ erg; Shigeyama et al. 2010; S10) 
in comparison with spherical SNe (S-SNe), 
it is highly likely that dwarf galaxies, like some of GRB hosts,
are strongly influenced by A-SN explosions.  
However, even the latest chemodynamical studies (e.g., Rahimi \& Kawata 2012)
have not investigated
chemical and dynamical influences of A-SN explosions on galaxies.

The purpose of this Letter is to show, for the first time,
that A-SN explosions can significantly influence
gas dynamics,  star formation histories, and chemical evolution
of dwarf galaxies by using our new chemodynamical simulations
of dwarf galaxy evolution.
In order to more clearly show the importance of A-SN feedback effects
in dwarf galaxy evolution,
we present the results of the following comparative models.
One is the ``ASN'' model in which SNe with their original masses
larger than $30 {\rm M}_{\odot}$ can become A-SNe whereas other
low-mass ($\le 30 {\rm M}_{\odot}$) can become S-SNe.
The other is the ``SSN'' model in which all SNe are S-SNe.
Although recent observations have shown that some massive
stars are fast-rotating (e.g., Groh et al. 2009),
it is observationally unclear what fraction of massive stars with masses larger
than $30 {\rm M}_{\odot}$ can become A-SNe. In the present paper,
we consider  two
extreme cases and thereby try to clearly point out the importance of
A-SNe in galaxy evolution.

\begin{figure}
\psfig{file=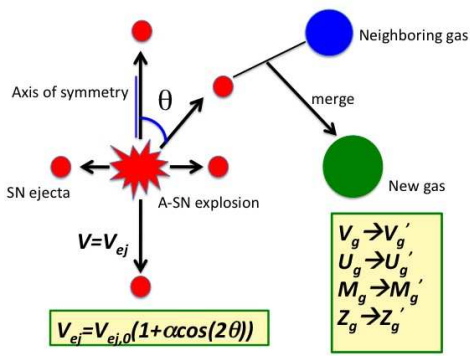,width=8.0cm}
\caption{
An illustrative explanation for the adopted new method to implement
feedback effects of A-SNe on galaxies. After the explosion of A-SN,
its gaseous ejecta can soon merge/mix with the neighboring gas (particle)
so that velocity ($V_{\rm g}$), internal energy ($U_{\rm g}$),
mass ($M_{\rm g}$), and metallicity ($Z_{\rm g}$) of the gas can change.
The ejection velocity ($V_{\rm ej}$) and chemical abundances 
of the ejecta depend on the angle ($\theta$) between the axis of symmetry
and the neighboring gas (particle).  The A3 model of S10 
(with $\alpha=7/9$) is adopted
for calculating $V_{\rm ej}$ and chemical yields.
}
\label{Figure. 1}
\end{figure}

\section{The model}
We investigate chemical evolution and star formation history
of an isolated  dwarf disk galaxy
by using our original chemodynamical code (``GRAPE-SPH'') that can be
run on a GPU cluster and the latest version of GRAPE
(GRavity PipE, GRAPE-DR) which is the special-purpose
computer for gravitational dynamics (Sugimoto et al. 1990).
We have revised our original GRAPE-SPH code (Bekki 2009)
by incorporating chemical yields both from S-SNe (Tsujimoto et al. 1995)
and from A-SNe (S10) so that we can investigate  not only
dynamical and hydrodynamical  influences 
of A-SN feedback on dwarf galaxies but also chemical enrichment
processes by A-SNe in a self-consistent way.
More details of the revised code (including code performance) will be
given in our forthcoming paper (Bekki 2012, in preparation)
and the results of one-zone chemical evolution models with ASN
for dwarf galaxies  will be discussed
in Bekki et al. (2012, B12).

\subsection{Dwarf disk galaxy}

A  dwarf is modeled as a fully self-gravitating system
and assumed to consist of a dark matter halo, a stellar disk,
and a gaseous one.
The density profile of the dark matter halo
with the total mass of $M_{\rm dm, dw}$ is represented by that proposed by
Salucci \& Burkert (2000):
\begin{equation}
{\rho}_{\rm dm}(r)=\frac{\rho_{\rm dm,0}}{(r+a_{\rm dm})(r^2+{a_{\rm dm}}^2)},
\end{equation}
where $\rho_{\rm dm,0}$ and $a_{\rm dm}$ are the central dark matter
density and the core (scale) radius, respectively.
Recent observational and numerical studies have shown that the adopted ``cored
dark matter'' halos are reasonable for describing dark matter distributions
in low-mass galaxies (e.g., Governato et al. 2010).
We adopt $M_{\rm dm, dw}=2.0 \times 10^9 {\rm M}_{\odot}$ and
$a_{\rm dm}=2.1$ kpc (i.e., $2.5 \times 10^8 {\rm M}_{\odot}$
within $a_{\rm  dm}$).

The stellar component of the dwarf is modeled as a  bulge-less stellar disk
with the total mass of $M_{\rm s,dw}$ and the size of 1.8 kpc.
The radial ($R$) and vertical ($Z$) density profiles of the stellar disk are
assumed to be proportional to $\exp (-R/R_{0}) $ with scale
length $R_{0}=0.36$ kpc  and to ${\rm sech}^2 (Z/Z_{0})$ with scale
length $Z_{0}=0.072$ kpc, respectively. 
We adopt $M_{\rm s, dw}=10^8 {\rm M}_{\odot}$ in this study.
In addition to the
rotational velocity caused by the gravitational field of disk
and dark halo components, the initial radial and azimuthal
velocity dispersions are assigned to the disc component according to
the epicyclic theory with Toomre's parameter $Q$ = 1.5.  The
vertical velocity dispersion at a given radius is set to be 0.5
times as large as the radial velocity dispersion at that point.
Dwarf irregular galaxies are observed 
to have gas disk extending to 
approximately twice the Holmberg radius ($R_{\rm H}$)
and  some of them have gas out to $(4-7)R_{\rm H}$ (e.g., Hunter 1997). 
We consider this observation and adopt an exponential gas disk
with the size of 7.2 kpc and the scale length
of 1.44 kpc. The total gas mass ($M_{\rm g, dw}$) is set to be
$10^8 {\rm M}_{\odot}$.

The radiative cooling processes dependent on gaseous metallicities
are properly included by using the MAPPING III code 
(Sutherland \& Dopita 1993). A gas particle is converted
into a new star if (i) the local dynamical time scale is shorter
than the sound crossing time scale (mimicking 
the Jeans instability)  and (ii) the local velocity
field is identified as being consistent with gravitationally collapsing
(i.e., div {\bf v}$<0$). 
We do not include additional model parameters for star formation,
such as the threshold gas density,
mainly because we try to more clearly demonstrate 
the differences in dwarf galaxy evolution between ASN and SSN models
in this study. 
The star formation efficiency is chosen such that
the efficiency can be proportional to $\rho_{\rm g}^{1.5}$
(to mimic the Schmidt law), where
$\rho_{\rm g}$ is the local gas density. The initial temperature is 
set to be $1.7 \times 10^2$ K. 
The gas disk is assumed to have (i) a radial metallicity gradient
with the central metallicity of [Fe/H]$=-1.32$ and a negative
radial gradient of $-0.2$ dex kpc$^{-1}$ and (ii) a SN-II like
enhanced [$\alpha$/Fe] ratio (e.g., [Mg/Fe]$\approx 0.4$).
Non-instantaneous recycling of
gaseous ejecta from type Ia and II SNs  and low-mass AGB stars are
properly considered for chemical enrichment processes associated with star
formation. 
The effects of UV background radiation, which could thermally heat up gas and
thus suppress star formation in dwarfs,  are not included.

\subsection{A new feedback scheme}

As shown in S10, chemical yields and velocities ($V_{\rm ej}$)
of SN ejecta in A-SNe
depend strongly on angles ($\theta$) between the axis of symmetry and
ejection directions. 
The axis of symmetry is set to be parallel to the $z$-axis of the
adopted coordinate system in which the $x$-$y$ plane coincides with
the disk  of a dwarf.
B12 have recently shown that
such $\theta$-dependences of chemical yields in A-SNe are important
for understanding chemical evolution of dwarf galaxies. We here adopt
the Model A3 of S10  in which the total explosion energy ($E_{\rm asn}$)
is assumed to be $10^{52}$ erg.  Owing to the $\theta$-dependences
of chemical yields and $V_{\rm ej}$,  we adopt different models
for feedback effects of S-SNe and A-SNe and the model for A-SNe is described
as follows (See Figure 1 for an illustrative explanation
of the A-SN feedback model).

A new stellar particle can release feedback energy of A-SNe to their neighboring
gas particles $1.2 \times 10^6$ yr after its formation. 
The half of the energy is used for the increase of the internal
energy of the gas (``thermal feedback'') and the other half
is used to change the momentum of the gas (``kinematic feedback'').
As a result of this, each of the neighboring gas particles changes
its mass ($M_{\rm g}$), internal energy ($U_{\rm g}$), 
velocity ($V_{\rm g}$), and metallicity ($Z_{\rm g}$). 
Guided by the results of 
2D hydrodynamical simulations for axisymmetric supernova explosions by
S10, we assume that the ejection velocity of A-SN is described as follows:
\begin{equation}
V_{\rm ej}(\theta) \propto  1+ (7/9) \cos(2\theta).
\end{equation}
Therefore $V_{\rm ej}$ is larger for SN ejecta close to jet 
(i.e., axis of symmetry).

We first estimate $\theta$ for each $j$-th neighboring gas particle
around $i$-the new stellar one and then change $M_{\rm g, \it j}$,
$U_{\rm g, \it j}$, $V_{\rm g, \it j}$, and $Z_{\rm g, \it j}$ according
to $\theta$-dependent chemical yields and feedback energy.
The new velocity of the $j$-th gas particle  after A-SN feedback effects
($V_{\rm g, \it j}^{'}$) 
is determined  by the following equation  (i.e., conservation of
linear momentum): 
\begin{equation}
(M_{\rm g, \it j}+m_{\rm ej}(\theta)) V_{\rm g, \it j}^{'} =
M_{\rm g, \it j}V_{\rm g, \it j}+m_{\rm ej}(\theta)V_{\rm ej}(\theta),
\end{equation}
where $m_{\rm ej}(\theta))$ is the mass of the A-SN ejecta at $\theta$.
Likewise,  the new thermal energy
($U_{\rm g, \it j}^{'}$) 
is determined by the following equation:
\begin{equation}
(M_{\rm g, \it j}+m_{\rm ej}(\theta)) U_{\rm g, \it j}^{'} =
M_{\rm g, \it j}U_{\rm g, \it j}+m_{\rm ej}(\theta)U_{\rm ej}(\theta),
\end{equation}
where $U_{\rm ej}(\theta)$ is the specific internal
energy  of the A-SN ejecta at $\theta$ and proportional
to $V_{\rm ej}^2(\theta)$. Therefore 
$E_{\rm asn}=
\int m_{\rm ej}(\theta) (0.5V^2_{\rm ej}(\theta) + U_{\rm ej}(\theta))
d\theta$.
The increase of chemical abundances in  the $k-$th heavy element 
due to chemical enrichment by A-SN of the $i$-th new stellar particle
is described as follows:
\begin{equation}
\Delta Z_{\rm g, \it j, k} =
m_{\rm ej}(\theta)A_{\rm ej, \it k}(\theta),
\end{equation}
where $A_{\rm ej, \it k}(\theta)$ is the chemical yield of A-SN
(normalized by mass).  
Each neighboring particle around a SN can receive the same amount
of metals.  Metal-diffusion between SPH particles is not
included, which could possibly overestimate the local abundance
inhomogeneity.

We adopt the same models of S-SN feedback effects as those used
in previous formation models of disk and elliptical galaxies
(e.g., Navarro \& White  1993; Kawata \& Gibson 2003). The total
energy of S-SN ($E_{\rm ssn}$) is set to be $10^{51}$ erg
and the half is used for thermal feedback and the other half
is used for kinematic feedback. 
We adopt the canonical Salpeter initial mass function (IMF)  with
the slope of $-2.35$ and the lower and upper mass cut-offs being
$0.1 {\rm M}_{\odot}$ and $50 {\rm M}_{\odot}$, respectively.
Therefore, the number ratio of SNe with masses of $[8-50] {\rm M}_{\odot}$
and those with masses larger than $30 {\rm M}_{\odot}$ is $\sim 10$. 
However, A-SNe can significantly influence evolution of dwarf galaxies 
owing to $E_{\rm asn} =10E_{\rm ssn}$.

\subsection{Simulation setup}
The total number of particles used for each component
(dark matter, stellar disk,
and gaseous one)  in a simulation is 200,000 (i.e., 600,000
in total). The gravitational softening length is fixed at
175pc for dark matter and 21pc for stellar and gaseous disks.
We investigate the star formation history and chemical 
evolution (in particular  Mg, Ca, and Fe) of dwarf galaxies
for 1.4 Gyr. Table 1 briefly summarizes the parameter values
used in the two comparative ASN and SSN models. 
In the following, $T$ in a simulation
represents the time that has elapsed since the simulation
started.

\begin{figure}
\psfig{file=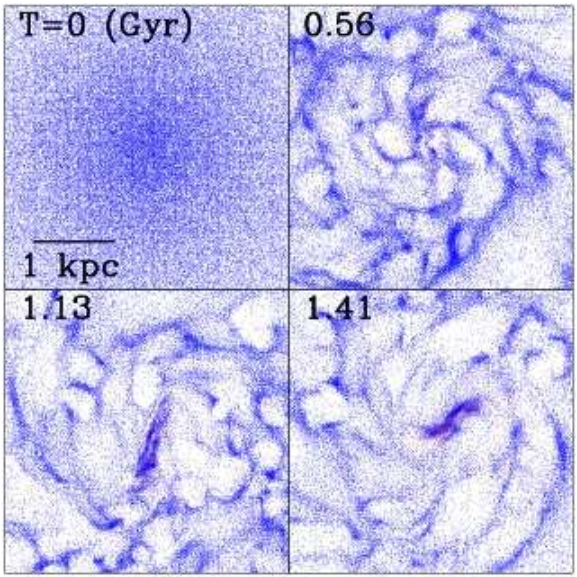,width=8.0cm}
\caption{
The time evolution of the distribution of gas (blue)
and new stars (red)  in the inner disk
projected onto the $x$-$y$ plane for the A-SN model.
The bar in the lower left corner of each panel measures 1 kpc
and the time $T$ is shown in the upper left corner.
}
\label{Figure. 2}
\end{figure}

\begin{figure}
\psfig{file=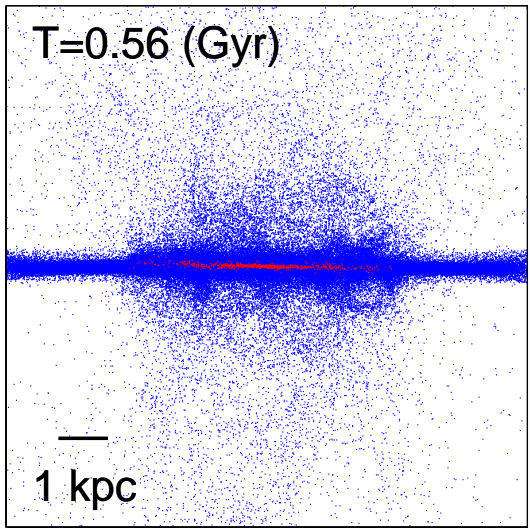,width=7.0cm}
\caption{
The distribution of gas (blue)
and new stars (red) 
projected onto the $x$-$z$ plane at $T=0.56$ Gyr for the ASN model.
}
\label{Figure. 3}
\end{figure}

\begin{figure}
\psfig{file=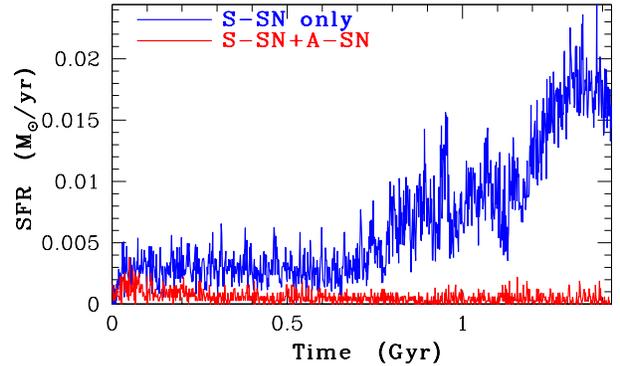,width=8.0cm}
\caption{
The time evolution of star formation rate for the SSN (blue) and
ASN (red) models.
}
\label{Figure. 4}
\end{figure}

\begin{figure}
\psfig{file=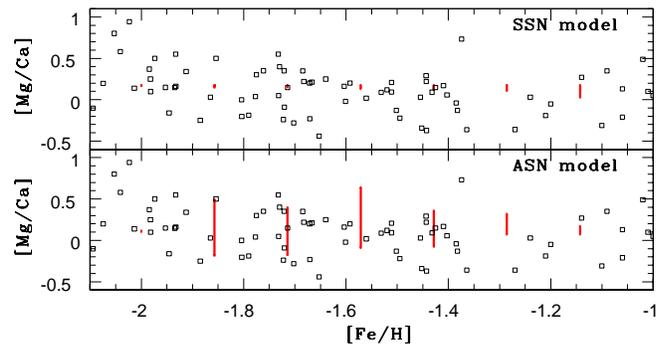,width=8.5cm}
\caption{
The simulated ranges of [Mg/Ca] (thick vertical red lines) for the
SSN (upper) and ASN (lower) models. The small open squares are observational
data for dwarfs in the Local Group from B12 which compiled the observational
data from different dwarf spheroidals (e.g., Koch 2009).
}
\label{Figure. 5}
\end{figure}

\section{Results}
Figure 2 shows the time evolution of the mass distribution
of the inner gas disk projected onto the $x$-$y$ plane in the ASN model.
As new stars form from high-density regions of the inner gas disk,
explosions of S-SNe and A-SNe can strongly disturb the surrounding local regions
through thermal and kinematic feedback effects ($T=0.56$ Gyr).
As a result of this,  the local regions with very low gas densities,
which can be identified as giant ``HI'' holes, can be formed in the inner
disk. These HI holes are connected with filamentary gaseous structures
with relatively high densities, where star formation can still continue.
A stellar bar can form by $T=0.84$ Gyr and a weak bar-like structure of gas
can be seen in the central region of the disk.
However, strong feedback effects of A-SN can prevent  the formation
of massive, high-density gaseous regions
in the center  and thus severely suppress the secondary
starburst there ($T=1.13$ and 1.41 Gyr).
The total mass of gas converted into new stars ($M_{\rm ns, dw}$) is 
only $9.7 \times 10^5 {\rm M}_{\odot}$ for the $\sim 1.4$ Gyr
evolution, which means that the stellar surface density can increase
only by $\sim 1$\%.

Figure 3 shows clearly that a significant fraction of the inner
gas disk  can be expelled
from the dwarf  owing to the strong SN feedback effects.
The total mass of gas with  $|z| > 10Z_0$ (=0.64 kpc), which can be regarded
as gas being expelled from the disk as ``stellar wind'',
is $1.8 \times 10^7 {\rm M}_{\odot}$. The rate of this stellar wind
is $3.2 \times 10^{-10} {\rm M}_{\odot}$ yr$^{-1}$, which is significantly
larger than that in the SSN model. About $1.2 \times 10^7 {\rm M}_{\odot}$
of the gas can be finally located outside the virial radius (15.6 kpc)
of the dwarf at $T=1.4$ Gyr so that the gas can not return back to
the original gas disk for further star formation.  The dwarf can finally
have a metal-poor 
gaseous halo with the total mass of $1.3 \times 10^7 {\rm M}_{\odot}$.

Figure 4 shows that global star formation in the ASN model
is more strongly suppressed in comparison with the SSN one owing
to the stronger SN feedback effects in the ASN model. The mean star formation
rate for 0 Gyr $\le T \le $ 0.56 Gyr (before stellar bar formation) is 
only $9.1 \times 10^{-4} M_{\odot}$ yr$^{-1}$, which is by a factor of $\sim 3$
smaller than that in the SSN model. The star formation rate can not 
rise even after the bar formation and ends up with the mean rate of
$6.8 \times 10^{-4} M_{\odot}$ yr$^{-1}$ for the 1.4 Gyr evolution. 
The difference in star formation rates between the two models
becomes even more remarkable 
after the bar formation ($T>0.8$ Gyr), because gas inflow due to
the bar can trigger a secondary starburst in the SSN model.
The star formation efficiency  
($\epsilon_{\rm sf}=M_{\rm ns, dw}/M_{\rm g, dw}$)
is estimated to be 0.0097 in the ASN model.

Figure 5 shows one of remarkable differences in chemical abundances
of new stars  at $T=1.4$ Gyr
between the ASN and SSN models in which  mean [Mg/Ca] of the ejecta for massive SNe is
$\sim 0.2$.
 The [Mg/Ca] abundance ratios
in the ASN (SSN) model range from $-0.18$ (0.03) to 0.64 (0.18): the large 
[Mg/Ca] scatter at a given [Fe/H] range can be clearly seen only in the ASN model.
The rather high [Mg/Ca] ($>0.2$) in the ASN model is due to chemical enrichment
by A-SNs (i.e., due to ejecta with high [Mg/Ca]).
As shown in Figure 5,
large scatter in [Mg/Ca] can be clearly seen in dwarf galaxies of
the Local Group. Therefore the present results imply that
one of possible explanations for the large [Mg/Ca] scatter in
the dwarfs is chemical pollution of interstellar medium  of the dwarfs by
ASN ejecta at their formation epochs.
As a result of very low star formation efficiency,  chemical enrichment
does not proceed efficiently so that the average [Fe/H] of new stars
can be kept low ($\sim -1.5$) in the ASN model.

\section{Discussion and conclusions}
We have first demonstrated that A-SNe can more strongly influence star formation
histories of dwarf galaxies than S-SNe owing to their larger amount of
feedback energy.
It should be however stressed
that the number fraction of A-SNe among all SNe with masses larger than
$30 {\rm M}_{\odot}$ ($f_{\rm asn}$) is assumed to be 1 in the present
A-SN model. This means that the present numerical study could overestimate
the feedback effects of A-SNe,  because it is observationally
unclear what a reasonable
value is for $f_{\rm asn}$ in dwarf galaxies.
Nagataki et al. (1997) showed that the observed light curve of 1987A in the
LMC is consistent with the predictions of their aspherical SN explosion models.
Recent observations on  
the presence of  fast-rotating massive stars
(Groh et al. 2009),
which are progenitors of A-SNe,
imply that A-SNe could  be ubiquitous.
These results imply that A-SNe could  be major  populations
among SNe and thus that future numerical simulations of galaxy formation and
evolution would need to include A-SNe self-consistently.

Bekki \& Tsujimoto (2012) failed to explain the observed low [Ca/Fe] 
($\sim -0.2$)
and high [Mg/Ca] ($\sim 0.3$) at [Fe/H]$>-0.5$ for the stars of the
LMC in their chemical evolution models with S-SNe only
and therefore suggested that A-SNe might play a role in chemical
enrichment processes of the LMC. B12 have recently shown that
the observed high [Mg/Ca] of the LMC and dwarfs can be reproduced well by
the chemical evolution models with A-SNe. 
As demonstrated in the present
chemodynamical simulations,  higher [Mg/Ca] can be achieved only
in the A-SN model.  These previous and present studies thus suggest
that the observed [Mg/Ca] can be an indicator of the long-term
influences of A-SNe  on chemical evolution of galaxies.

The present study predicts that if dwarf disk galaxies continue to experience
A-SN feedback effects, they can end up with galaxies with faint luminosities and 
low surface brightness (LSB) owing to very low conversion efficiency of gas
into new stars. This prediction implies that (i) the origin of  LSB galaxies
could be closely associated with significantly stronger influences
of A-SN feedback effects (i.e., higher $f_{\rm asn}$)
and (ii) LSB might well have unique chemical abundance patterns
(e.g., high [Mg/Ca]).
The present results also suggest that $f_{\rm asn}$ 
is a  key parameter that can control how the SN feedback can strongly
suppress star formation and thereby determine the  structures (e.g., surface mass
densities) of dwarf galaxies.
For example,
if $f_{\rm asn}$ is significantly higher ($>0.1$) in a dwarf at its formation epoch,
then the dwarf could finally become
a very faint LSB galaxy like ultra-faint dwarfs.

Investigation of A-SN feedback effects on galaxies have just
started in the present study and the better and realistic
ways to implement feedback effects 
from different SNe (e.g., how to include kinematic effects etc) 
need to be carefully investigated.
It is our future work to investigate how galaxy formation processes
through hierarchical merging of subgalacitc clumps can change if
more energetic A-SN feedback effects are properly included
in our future chemodynamical simulations.

\section{Acknowledgment}
We are  grateful to the referee  for  constructive and
useful comments that improved this paper.
Numerical simulations  reported here were carried out on 
the GPU cluster 
gSTAR kindly made available by the Center for Astrophysics and Supercomputing
in the Swinburne University.

\end{document}